\documentclass[12pt]{iopart}
\usepackage[dvips]{graphicx}

\newcommand{\tpulse}{\tau}
\newcommand{\tperiod}{\tau}
\newcommand{\temp}{T}  
\def\VUV{{\scriptsize VUV}}
\def\XUV{{\scriptsize XUV}}
\def\vuv{{\small VUV}}
\def\xuv{{\small XUV}}
\def\nir{{\small NIR}}

\begin{document}

\title[Non-equilibrium plasma dynamics on the attosecond 
timescale]{Tracing non-equilibrium plasma dynamics on the
  attosecond timescale in small clusters}

\author{Ulf Saalmann, Ionu\c{t} Georgescu and Jan M.~Rost}  
\address{Max Planck Institute for the Physics of Complex
  Systems\\ N\"othnitzer Stra{\ss}e 38, 01187 Dresden, Germany} 
\ead{us@pks.mpg.de}
\begin{abstract}
  It is shown that the energy absorption of a rare-gas cluster
  from a vacuum-ultraviolet (\VUV) pulse can be traced with
  time-delayed extreme-ultraviolet (\XUV) attosecond probe pulses
  by measuring the kinetic energy of the electrons detached by
  the probe pulse.  
  By means of this scheme we demonstrate, that for pump pulses
  as short as one femtosecond, the charging of the cluster
  proceeds during the formation of an electronic nano-plasma
  inside the cluster.  
  Using moderate harmonics for the {\VUV} and high harmonics for
  the {\XUV} pulse from the same near-infrared  laser source, this
  scheme with well defined time delays between pump and probe
  pulses should be experimentally realizable.  
  Going to even shorter pulse durations we predict that pump
  and probe pulses of about 250\,attoseconds can induce and
  monitor non-equilibrium dynamics of the nano-plasma.
\end{abstract}


\maketitle

\section{Introduction}
The development of ultrashort attosecond laser pulses is taking
breathtaking development.  With isolated pulses already technically
possible \cite{sabe+06}  attosecond pulses or pulse trains are
presently used in combination with a strong near-infrared ({\nir})
pulse from which the attosecond pulses were generated in the
first place.  
Although the strong {\nir} pulse modifies in many cases the target,
it conveniently provides a clock which indicates at which time
$t$ the atto pulse acted on an electron, since the final momentum
$\vec{p}_\mathrm{final}=\vec{p}_\mathrm{excited} +
e\vec{{\cal A}}(t)$ of an ionized electron contains as drift momentum
the vector potential $\vec{{\cal A}}(t)$ of the {\nir} field at time
$t$ and is therefore streaked by the {\nir} pulse
\cite{drhe+02,kigo+04}.

We have proposed a different pump-probe scheme: A {\vuv} pulse
(about 100\,fs duration and 20\,eV frequency) excites a rare-gas
cluster, and the evolution of the electron dynamics is traced by
a time-delayed attosecond pulse.  
While such a combination promises to deliver
insight into fast dissipative multi-electron dynamics in the cluster
\cite{gesa+07}, it is still experimentally out of reach.  However,
one can generate the {\vuv} pulse from harmonics of a strong {\nir} pulse
and combine it with an atto pulse.  The prize to pay is that the {\vuv}
pulse will be relatively short, namely a few femtoseconds which is
almost a factor 100 shorter than before \cite{gesa+07}.  Here, we explore
this new scenario which, moreover, gives rise to new phenomena in the
cluster dynamics as will be detailed below.

After introducing our approach in Sect.\,2, we analyze in Sect.\,3
the dynamics of small argon clusters with 13 and 55 atoms under a {\vuv}
pulse of a few femtoseconds and an attosecond probe pulse, focusing on
mapping out the charging of the cluster which happens dominantly
during the rising part of the {\vuv} pulse.  This scenario extrapolates
the one studied before \cite{gesa+07} to shorter excitation times.  We
find an interesting new phenomenon in a small time window at the
crossover between instant and delayed ionization.  To understand this
phenomenon we replace the pump pulse in Sect.\,4 with a sudden
excitation of electrons.  It turns out that their subsequent dynamics
is that of a strongly-coupled non-equilibrium plasma
\cite{popa+05} which oscillates 
between kinetic and potential energy with the plasma frequency, indeed
explaining the crossover found in Sect.\,3 as the reminder of this
non-equilibrium dynamics when excited and probed with relatively long
pulses.  Consequently, we demonstrate in Sect.\,5 that one can indeed
observe this non-equilibrium plasma dynamics if attosecond pump and
probe pulses are used, by tracing the potential energy in form of the
charging of the cluster, as described before.  We summarize our
results in Sect.\,6.

\section{Theoretical method}
In order to simulate the dynamics of electrons and cluster 
ions upon excitation by  {\vuv} laser pulses   
we have used a hybrid approach which combines quantum-mechanical 
absorption rates for bound electrons and classical propagation
of the photo-ionized electrons.
The approach is similar to those successfully applied to cluster
dynamics driven by {\nir} laser pulses \cite{sasi+06}.
Details of the adaption to the {\vuv} regime have been published 
elsewhere \cite{gesa+07a}. 
Therefore we will only briefly review the main steps and 
assumptions here.
Bound electrons may absorb photons according  to known 
photo-absorption rates \cite{co81}, i.\,e., there is a certain
probability for absorption within a given time step.
Due to this statistical feature one has to average over an
ensemble of realizations, where each realization is a deterministic 
ionization sequence and the ensemble, containing typically about 100 
realizations, is in accordance with the absorption rate.
After their ``creation'' the electrons are propagated classically in the
field of the laser and all previously generated electrons and ions.
In contrast to photo-absorption of isolated atoms, electrons are
not necessarily free but may be trapped in the cluster potential,
i.\,e., by the attractive Coulomb potential of neighbouring
ions.
These electrons are called quasi-free electrons.
Whereas the classical propagation is straight forward, it is
challenging to properly describe the impact of the electrons and
ions on further ionization of bound electrons. 
We have developed a method for calculating the photo-ionization
rates in clusters which takes into account
localization of  electrons around a particular ion  \cite{gesa+07a}.
This ``classical recombination'' is of minor importance for the
short pulses studied here but becomes relevant for longer pulses 
\cite{gesa+07} as produced by free-electron-laser sources.
 
\section{Observation of transient   cluster charging}\label{sec:transient}

We will study the transient charging of small clusters using a
pump-probe scenario proposed recently~\cite{gesa+07}.
The clusters are excited by a  {\vuv} pump pulse which lasts\,---\,in 
contrast to our previous study\,---\,only for a few
femtoseconds and probed with time-delayed attosecond {\xuv} pulses.
The instantaneous charge at time $t$ is imprinted in the final
kinetic energy of the electron released during the attosecond
pulse which is shifted with respect to the peak of the pump
pulse by a time delay $t$.
Slower electrons indicate higher charges.
In order to obtain absolute values we compare the energy
reduction to the series of ionization potentials of the
subsequent charge states of argon.
The parameters chosen for the pump and probe pulses are
motivated by their experimental availability~\cite{kr07}.  
Both pulses are most easily obtained from filtering of
(different) harmonics from a driving {\nir} laser. 
Besides a flexible duration of the pump pulse, most importantly,
this enables an accurate setting of the delay of the probing {\xuv}
pulse.  
The intensity of the probe pulse is of minor importance.
On average less than one electron is photo-ionized per pulse due
to the low cross sections for {\xuv} frequencies. 
A lower intensity can be compensated by a higher repetition
rate.

\begin{figure}
\centerline{\includegraphics[scale=0.7]{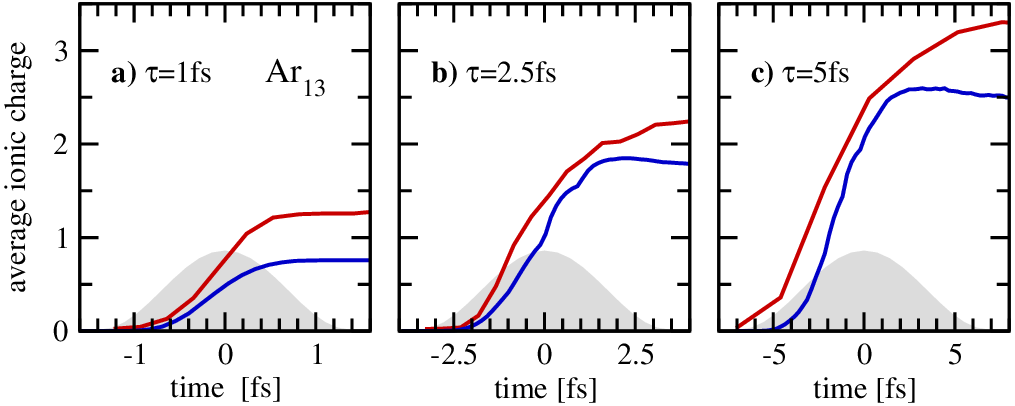}}
\caption{Average charge of ions in an Ar$_{13}$ induced by laser
  pulses (gray shaded areas) with three different pulse lengths
  $\tpulse=1$\,fs, 2.5\,fs, and 5\,fs and the central frequency
  $\hbar\omega=20$\,eV and the same intensity
  $I=7\times10^{13}$\,W/cm$^2$.  
  We compare the ionic charge (blue lines) due to to
  photo-ionization of the pump pulse with the charge probed by
  the kinetic of the delayed attosecond pulse (red lines), see
  text for details.
}
\label{fig:inner13}
\end{figure}
\begin{figure}[b]
\centerline{\includegraphics[scale=0.7]{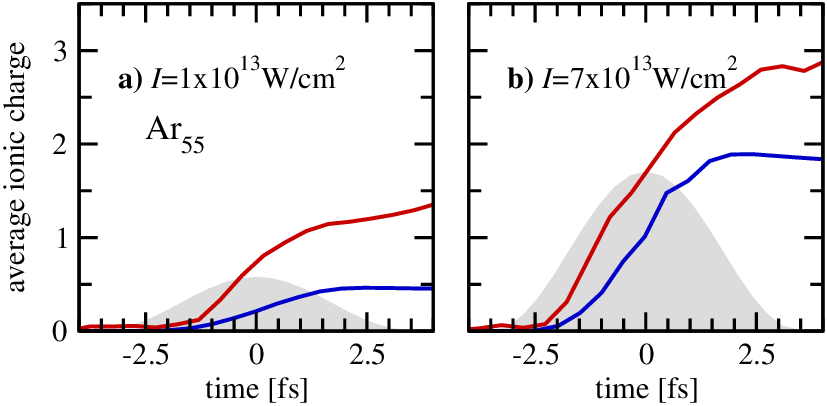}}
\caption{Same quantities as in Fig.\,\ref{fig:inner13}, but for an
  Ar$_{55}$ cluster and laser pulses of the same duration
  $\tpulse=2.5$\,fs and but two different intensities
  $I=1\times10^{13}$\,W/cm$^2$ and $7\times10^{13}$\,W/cm$^2$.
}
\label{fig:inner55}
\end{figure}

%
Figure \ref{fig:inner13} and \ref{fig:inner55} show the charging
for Ar$_{13}$ and Ar$_{55}$ clusters, respectively.
In both cases the pump frequency was $\hbar\omega=20$\,eV.
For the smaller cluster we kept the intensity $I$ constant and
changed the pulse duration $\tpulse$;
for the larger cluster we did it \textsl{vice versa}.
Photo-absorption due to the pump pulses leads to excitation of
electrons either directly into the continuum or\,---\,due to the
increasing space charge\,---\,into the cluster.
In either of the cases the electron is lost from the
mother ion and the ionic charge  increases.
These charges, averaged over all cluster ions, are shown by the
blue lines in Figs.\,\ref{fig:inner13} and \ref{fig:inner55}.
The maximum of the ionization rates (strongest increase) shifts
to earlier times if the pulse is made longer (as from
Figs.\,\ref{fig:inner13}a to c) or if the intensity is increased
(as from Figs.\,\ref{fig:inner55}a to b).
This is due to the depletion of the atoms in the cluster. It
becomes relevant for average ionic charges larger than one,
which is the case in Figs.\,\ref{fig:inner13}b, c, and
\ref{fig:inner55}b. 
For the ``long'' pulse with $\tpulse=5$\,fs one observes localization
or recombination of electrons which results in a decreased charge
at the falling edge of the pulse, i.\,e., for times $t>2.5$\,fs.
During the pulse we do not find noticeable localization.

The ionic charge as obtained from the attosecond probe pulses
($\hbar\omega=150$\,eV, $I=10^{15}$\,W/cm$^2$,
$\tpulse=500$\,as) is shown by red lines in the
Figs.\,\ref{fig:inner13} and \ref{fig:inner55}. 
In this case the horizontal axis corresponds to the time delay
of the probe pulses.  
Although the probing does not resemble the charging process
perfectly for all cases, it performs particularly well when the 
charging is strongest, namely slightly before the pulse maximum.
There are basically three reasons for the apparent deviations: 
i) The space charge reduces the kinetic energy of the probing
electrons and results in an overestimation of the ion charges.
This effect is more important for Ar$_{55}$ as compared to
Ar$_{13}$ because of the larger cluster charge.  
ii) Evaporation of quasi-free electrons increases the space
charge as well.  
This effect becomes relevant only late in time, in our examples,
after the pump pulse is over, see the deviation of both
quantities in Fig.\,\ref{fig:inner13}c.  
iii) For either very short (Fig.\,\ref{fig:inner13}a) or very
weak (Fig.\,\ref{fig:inner55}a) pulses both quantities show an
(almost linearly) increasing divergence already at early times.
Note the difference to the other cases, where both curves are
largely parallel during the initial charging.  
The increasing divergence is due to the low number of electrons
captured by the cluster potential.  
Since the cluster does not host a plasma which would allow for
screening of the ionic charges, the increasing space charge is
directly imprinted in the probing signal.

\begin{figure}[b]
\centerline{\includegraphics[scale=0.7]{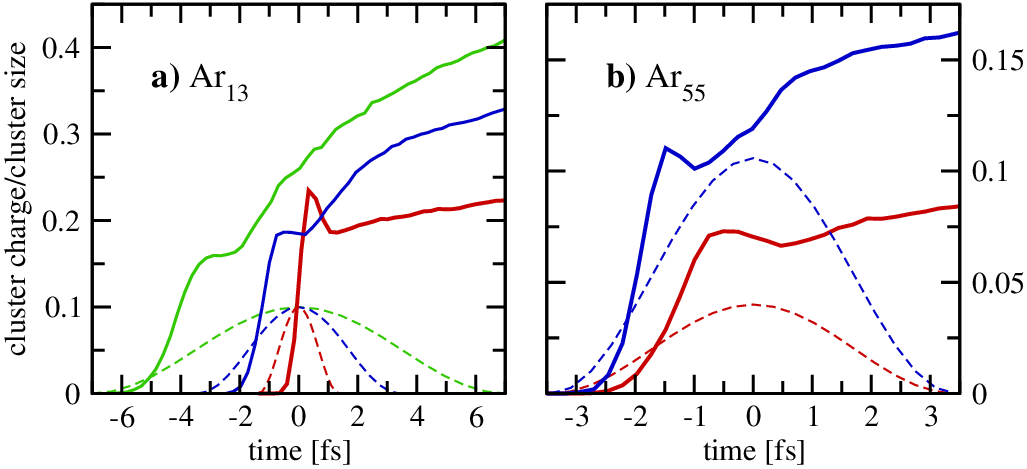}}
\caption{Total charge of the cluster, i.\,e., number of
  electrons with positive energy, divided by the cluster
  size. Left panel: cluster and laser pulse parameters as in
  Fig.\,\ref{fig:inner13}, right panel: as in
  Fig.\,\ref{fig:inner55}.
  The envelopes of the electric fields for the corresponding
  pump pulses are shown by dashed lines. 
}
\label{fig:outer}
\end{figure}%
Since the total cluster charge influences the probing signal, it
is presented in Fig.\,\ref{fig:outer} as a function of time for the
cases considered above.
It is defined by the number of electrons with positive energy
$E>0$.
The energy of the $j$th electron (mass $m$, charge $-e$) 
with velocity $\vec{v}_j$ is given by 
\begin{equation}
  \label{eq:enel}
  E_j=\frac{m\vec{v}_j{}^2}{2}-\sum_{i(\ne j)}^{\mathrm{all}}
  \frac{q_i\,e^2}{|\vec{r}_i-\vec{r}_j|},
\end{equation}
whereby the sum runs over all (except the electrons itself)
charged particles with charge $q_i\,e$ at the position $\vec{r}_i$.
In order to make Ar$_{13}$ and Ar$_{55}$ comparable the cluster
charge is normalized to the  number of cluster atoms.
In all cases the charging of the cluster shows clearly two
phases, namely fast direct charging early in the pulse and slow
evaporation towards the end of the pulse and later on.
Whereas for the longer pulses and Ar$_{13}$, cf.\ green and blue
lines in Fig.\,\ref{fig:outer}a, these phases are separated by a
plateau, in the other cases the cluster charge even drops after a
maximum before evaporation sets in. 
Comparison with Figs.\,\ref{fig:inner13} and \ref{fig:inner55}
reveals that this ``overshooting'' is connected with a high
photo-absorption rate, i.\,e., a large number of atoms becomes
ionized within short time interval.
This is either due to a quick rising\footnote{The ionization
  rate in panel a of Fig.\,\ref{fig:inner13} is considerably
  larger than in panel c, which is somehow hidden by the
  different time scales.}   
of the laser intensity, as for the 1-fs-pulse, or due a large
number of available atoms, as for Ar$_{55}$.
The plateau or even the overshooting of the charge at the
crossover from direct ionization to evaporation is most easily
understood if one considers the limit of instant ionization of
the electrons.

\section{Instantaneous cluster ionization:
Formation,\\ equilibration and relaxation of a nano-plasma} 

\begin{figure}
  \centerline{\includegraphics[scale=0.7]{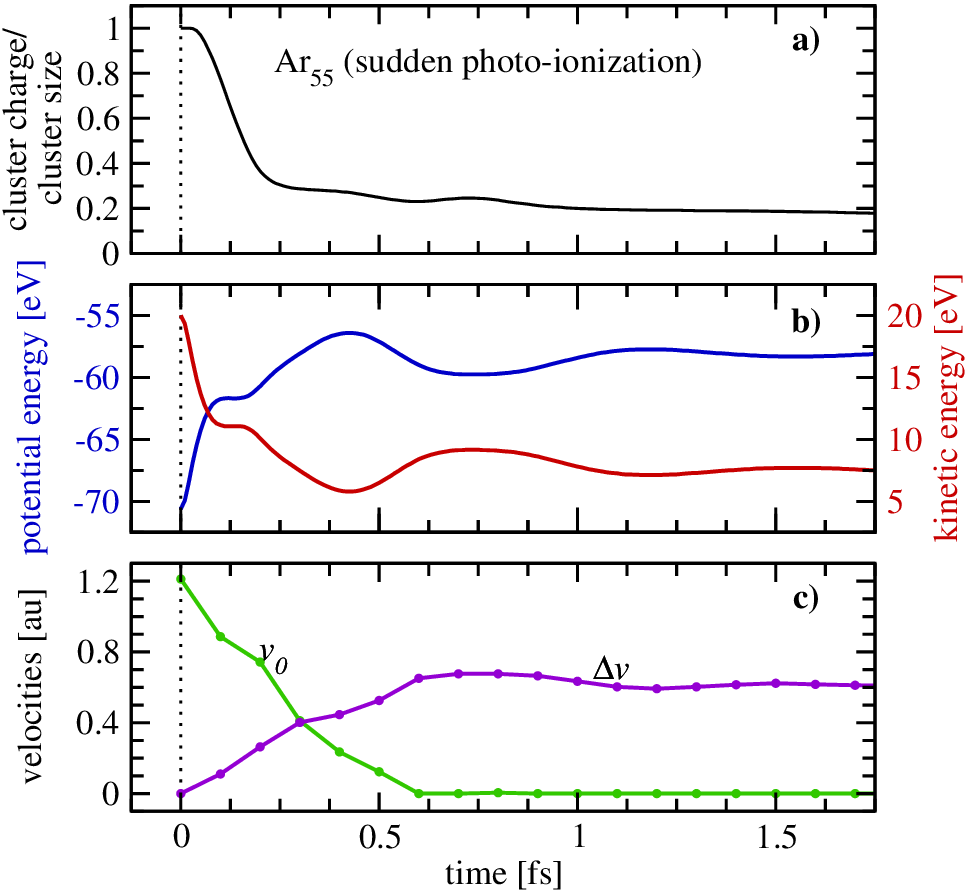}}
  \caption{Time evolution after sudden single-photo-ionization of
    all atoms of an  Ar$_{55}$ cluster.
    From top to bottom:
    \textbf{a)}  total charge of the cluster divided by the cluster
    size, 
    \textbf{b)} potential (blue, right axis) and kinetic
    (red, left axis) energy of quasi-free electrons, i.\,e.\
    electrons with negative  energy,
    \textbf{c)} fit parameter for the velocity distribution of
    Eq.\,(\ref{eq:vfit}). 
  }
  \label{fig:model}
\end{figure}

Modelling instant ionization we assume for simplicity that all
electrons absorb simul\-taneously one photon.  Figure~\ref{fig:model}
shows the results for an Ar$_{55}$ cluster where one electron per
atom was released at time $t=0$ with an excess energy $E_0=4.24$\,eV,
according to a photon frequency $\hbar\omega=20$\,eV and the ionization
potential of neutral argon $E_\mathrm{IP}=15.76$\,eV; from that time
on they are propagated classically.  At the time of creation all
electrons have positive energy and the cluster charge is equal to the
cluster size as seen in Fig.\,\ref{fig:model}a.  This corresponds to
the temporary enhancements seen in Fig.\,\ref{fig:outer} and
discussed before.  The value drops down to about 0.2 similar to the
values observed in the realistic calculations for Ar$_{13}$ and
Ar$_{55}$.  This drop is easily understood considering that the
Coulomb energy of 55 singly-charged ions is larger than 3\,keV. The
total excess energy of the 55 electrons is, however, only about
233\,eV. Thus for energetic reasons the majority of the electrons
cannot leave the cluster volume.

More insight regarding the electron dynamics is gained by taking a
closer look on the electron energies.  Figure~\ref{fig:model}b shows
separately the potential and kinetic energy of all quasi-free
electrons.  Their sum is approximately constant, violated only
slightly through electrons which leave the cluster.  This violation is
small because, firstly, the number of electrons which leave the
cluster during this time is negligible and, secondly, those which are
lost carry only a very small amount of energy.  Apparently there is an
oscillatory exchange of potential and kinetic energy.
The period of this oscillation is
$\tperiod_\mathrm{osc}\approx0.79$\,fs, which agrees 
with plasma period
$\tperiod_\mathrm{pl}=2\pi/\omega_\mathrm{pl}
=\sqrt{\pi m/\varrho e^2}=0.77$\,fs
according to the atomic density $\varrho$ of argon at equilibrium
structure\footnote{The density is calculated for 55 atoms and a
cluster radius of 8.5\,\AA. Atomic motion due to Coulomb explosion can
be neglected on the considered time scale of a few femtoseconds.}.
Obviously those electrons which could not leave the cluster execute
plasma oscillations.

That a plasma has formed on this short time scale may be surprising
but can be seen in Fig.\,\ref{fig:model}c.  For selected times, shown
by dots in Fig.\,\ref{fig:model}c, we have calculated the velocity
distribution of the quasi-free electrons.  This distribution is fitted
well for all times by the following function \cite{maro+57}
\begin{equation}
  \label{eq:vfit}
  f_{v_0,\Delta v}(v)=C\cdot v^2\cdot\exp
  \left(-\frac{(v{-}v_0)^2}{\Delta v^2}\right)
\end{equation}
with $C$ an irrelevant normalization constant.  The distribution of
Eq.\,(\ref{eq:vfit}) contains both limits: monoenergetic electrons
(with energy $E_0$) which corresponds to $v_0=\sqrt{2E_0/m}$ and $\Delta
v\to0$ and an electron plasma at equilibrium (with temperature $\temp$)
which corresponds to $v_0=0$ and $\Delta v=\sqrt{2k\temp/m}$.  Hence, the
ratio $v_0/\Delta v$ characterizes the degree of relaxation to
equilibrium as a function of time.  Starting with an infinitely large
value at time $t=0$\,fs the ratio vanishes around $t=0.6$\,fs.  Thus,
the relaxation of the electronic distribution function takes about
$\tperiod_{\mathrm{rel}}=0.6$\,fs which is of the same order as the built-up
of electronic correlations characterized by the plasma period
$\tperiod_\mathrm{corr}=0.77$\,fs.  It is known from one component plasmas
\cite{zw99} as well as from ultracold plasmas \cite{kili+01}
that in this situation the plasma temperature undergoes oscillations
about its equilibrium value if the plasma is strongly coupled,
i.\,e., if the Coulomb-coupling parameter
\begin{equation}
  \label{eq:ccpa}
  \Gamma=\frac{E_\mathrm{pot}}{E_\mathrm{kin}}>1,
\end{equation}
with $E_\mathrm{pot}$ and $E_\mathrm{kin}$ the potential and
kinetic energy of the quasi-free electrons, respectively.
For the system studied here we find $\Gamma\approx6$.
The oscillations are damped out quickly in about two
plasma periods which is typical for an inhomogeneous plasma 
\cite{popa+05}.
Discussing  kinetic and potential energy of the electrons as a 
function of time assumes that we have an ideal ``probe'' of infinite 
time-resolution to record these energies. In the next section we 
discuss how close one can come to this ideal with attosecond 
pulses.

\section{Creating and monitoring of non-equilibrium
plasmas\\ in clusters with attosecond pump-probe pulses}

With today's attosecond technology it is possible to come very close
to the sudden excitation assumed in the previous section.
If both, the pump \emph{and} the probe pulse, have  
sub-femtosecond length, the non-equilibrium plasma oscillations can 
be observed by monitoring the time-dependent charging of the 
cluster as described in Sect.\,\ref{sec:transient}.
We have chosen $\tpulse=250$\,as in order to ensure
that the creation and  probing of the plasma is fast on its
time scale given by the plasma period of 770 as (see previous 
section). 
Otherwise the scenario is equivalent\footnote{In
  contrast to the femtosecond pulses of Sect.\,\ref{sec:transient} one
  has to consider the large bandwidth of the attosecond pulse, which
  is taken into account by initiating the classical electron motion 
  after photo-absorption with the
  corresponding energy spread.} to that in Sect.\,\ref{sec:transient}.
We measure the kinetic energy of electrons kicked out by the probe
pulse ($\hbar\omega=150$\,eV) as a function of the time delay. 
Thereby, we obtain  the time-dependent charge which is close to the 
charge of the cluster if the pump pulse is short  as discussed in
Sect.\,\ref{sec:transient}.

\begin{figure}[b]
\centerline{\includegraphics[scale=0.7]{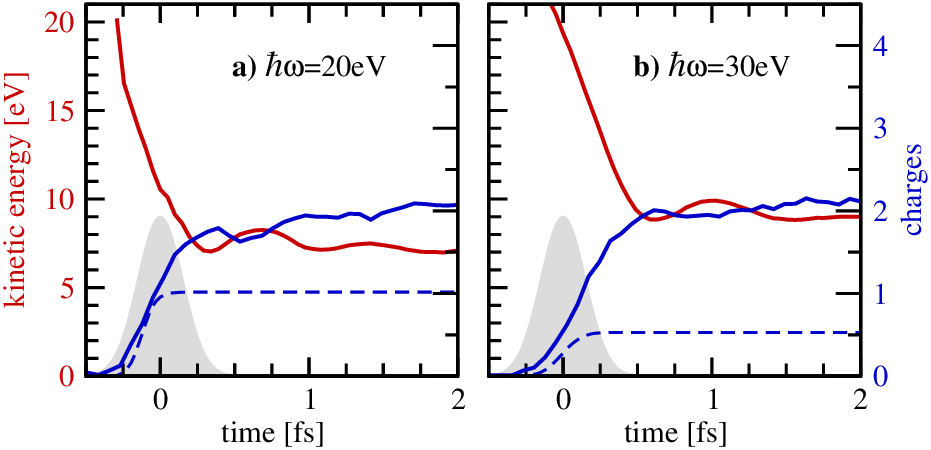}}
\caption{Time evolution after pumping of an  Ar$_{55}$ cluster
  by an attosecond {\vuv} pulse (gray shaded areas) with
  duration $\tpulse=250$\,as, intensity $I=5\times10^{14}$\,W/cm$^2$
  and \textbf{a)} $\hbar\omega=20$\,eV or
  \textbf{b)} $\hbar\omega=30$\,eV, respectively 
  The average kinetic energy (red lines, left axis) is shown
  along with the probed charge (solid blue lines, right axis).
  For completeness we show the ionic charges (dashed blue lines,
  right axis) which stay constant after the pulse.
}
  \label{fig:atto}
\end{figure}%
Figure~\ref{fig:atto} shows the probed charge (blue solid lines)
for two pump laser frequencies $\hbar\omega=20$\,eV and 30\,eV.
Evidence for plasma oscillations are given by the calculated
kinetic energy of the electrons (red solid lines).
Oscillations become apparent after the initial energy drop
during which the plasma formation occurs. 
They are more pronounced for the lower pump
frequency (Fig.\,\ref{fig:atto}a) where on average one electron
per atom is photo-ionized (blue dashed line).
In the other case (Fig.\,\ref{fig:atto}b), with a lower electron
density and longer plasma period, only one oscillation is
visible.
Similar to this \emph{theoretical} quantity one observes a
stronger contrast in the \emph{measurable} probed charge for
$\hbar\omega=20$\,eV than for $\hbar\omega=30$\,eV. 
These oscillation are solely due to oscillating charge
distribution; the ionic charges (blue dashed lines) remain
constant after the pulse is over. 

Although the attosecond pulse generated nano-plasma in the
cluster differs by orders of magnitude in density, temperature
and absolute number particles from an ultracold
``micro''--plasma \cite{kipa+07}, we find very similar behavior
for both finite plasmas, namely a fast decay of the plasma
oscillations and more pronounced plasma oscillations for a
larger  Coulomb-coupling parameter $\Gamma$.
In our case this is realized for 20\,eV excitation frequency as
compared to 30\,eV, where for the former the density is higher 
(cf.\ blue dashed lines in Fig.\,\ref{fig:atto})  
and the temperature is lower 
(cf.\ red lines in Fig.\,\ref{fig:atto}), 
hence the plasma is more strongly coupled.
These observations are in accordance with the predictions from
extendend plasmas \cite{zw99}.

\section{Summary}
We have demonstrated that an attosecond pump-probe scheme opens
the way to study non-equilibrium plasma phenomena in
nano-plasmas created from finite systems such as rare-gas
clusters. 
Initiated by a {\vuv} pump pulse, the {\xuv} probe pulse can map
out energy absorption of the finite multi-electron
system\,---\,represented in our case by a small rare-gas
cluster\,---\,\emph{during} the creation of the electronic
nano-plasma.  
Alternatively, the probe pulse can be used to trace the
non-equilibrium dynamics of the plasma including its relaxation
\emph{after} its creation. 
In both cases, ultrafast photo-ionization of the cluster ions
triggered by the probe pulse is used to determine the charging
of the cluster which provides directly the information on energy
absorption in the first case and on plasma oscillations
indicative of non-equilibrium dynamics in the second case. 

The fast time scale for the formation of correlation as well as the
plasma oscillations indicate the formation of a strongly-coupled
plasma which has been built up due to the high density of the cluster
ions in combination with the relatively low energy of the electrons
\cite{raju+06}.  Hence, the scheme proposed offers a route to make
strongly-coupled plasmas in clusters experimentally accessible, which
would boost our knowledge on dynamics in finite correlated
multi-electron systems.  These aspects will be investigated in more
detail in future work \cite{gesa+07d}.

\end{document}